# Biosorption of Cr(III), Cr(VI), Cu(II) ions by intact cells of *Spirulina platensis*

E. Gelagutashvili, N.Bagdavadze, A.Rcheulishvili

Iv. Javakhishvili Tbilisi State University
E. Andronikashvili Institute of Physics
0177, 6, Tamarashvili St.,
Tbilisi, Georgia
E.mail:eterige@gmail.com

## Abstract

The absorption characteristics of Cr(III), Cr(VI), Cu(II) ions on intact living cells *Spirulina platensis* (pH9.6) were studied by using a UV-VIS spectrophotometer; also biosorption of these ions with cyanobacteria *Spirulina platensis* were studied using equilibrium dialysis and atomic-absorption analysis.

It was shown, that the absorption intensity of *Spirulina platensis* decreases, when Cr(III), Cr(VI), Cu(II) ions are added. Significant difference between the absorption intensity for Cu(II) – *Spirulina platensis* and Cr(VI)_, Cr(III)_ *Spirulina platensis* were observed. Cu(II) was more effectively adsorbed by cyanobacterium than Cr(VI)_ and Cr(III).

The biosorption constants for Cu(II)_ *Spirulina platensis* is higher than for Cr(VI) _ and Cr(III) _*Spirulina platensis* 2.69 and 10.15- fold respectively.

The search for new technologies for the removal of heavy metals from wastewaters has been directed towards biosorption. Different types of microorganisms, have been shown to be efficient for metal biosorption.

Copper is widely present in the environment. Copper is present in various industrial effluents. it can contaminate drinking water in the case of corrosion of household copper pipes. It has been found that marine life is damaging due to high copper concentration in water [1]. Copper removal from wastewaters has a very importance.

Chromium exists in several oxidation states, however most that can be found in the environment are the trivalent Cr(Ill) and hexavalent Cr(VI) species. Cr (III) is reported to be less toxic than Cr (VI). Cr(VI) is carcinogenic and mutagenic for all living organisms. Since Cr(III) is insoluble at neutral and high pH, chemical reduction of Cr(VI) to Cr(III) followed by precipitation as $Cr(OH)_3$ has been long applied as a method for decontaminating waters from chromate [2].

Many researchers have recently demonstrated the possibility of removing heavy metals. In work[3], adsorption of chromium (VI) on fresh and spent *Spirulina platensis* algal cells was studied. Physico-chemical parameters are reported to have a strong influence on adsorption of metal ions, especially when using living biomass [4]. The mechanism by which binding of ions occurs is not yet completely understood.

In this paper the absorption characteristics of Cr(III), Cr(VI), Cu(II) ions on intact living cells *Spirulina platensis* were studied by using a UV-VIS spectrophotometer, also biosorption of these ions with cyanobacteria *Spirulina platensis* were studied using equilibrium dialysis and atomic-absorption analysis.

## Materials and Methods

Analytical grade reagents were used: $K_2CrO_4$, $CrCl_3$, $CuCl_2$. In these experiments were used *Spirulina platensis* living cells. *Spirulina platensis* IPPAS B-256 strain was cultivated in a standard Zaroukh [5] alkaline water-salt medium at 34ºC, illumination ~ 5000 lux, at constant mixing [6]. The cultivation of the *S. platensis* cells was conducted for 6 days. The intact cells suspendion of *Spirulina platensis* (pH 9.6) in medium was taken for scanning the absorption spectra from 380-850 nm by using a UV-VIS spectrophotometer.

The experiments of dialysis were carried out in 5ml cylindrical vessels made of organic glass. A cellophane membrane of 30µm width (type - "Visking" manufacturer - "serva") was used as a partition. The duration of dialysis was 72 hours. The experiments were carried out at $23^0C$ temperature. Errors between experimental data and mean values never exceeded ±13%.

In all mentioned cases, the concentration of *Spirulina platensis* was 1.6 mg/ml. It was determined by the instrumentality of [7, 8] data. The metal concentration ranged from $10^{-3}$ to $10^{-5}$ M were prepared in deionized water as series of standards. It was measured after dialysis by atomic absorption analysis „Analyst-900'' (Perkin Elmer) at the wavelength of $\lambda_{Cu}$=324,75 nm and $\lambda_{Cr}$=357.9 nm. Each value was determined as an average of three independent estimated values ± the standard deviation. The isotherm data were characterized by the Freundlich [9] equation. The parameters appearing in equation were estimated by linear regression, in the log versus log $C_b$ vs log$C_{total}$, where $C_b$ is the ion concentration connected with cyanobacteria for all cases, and $C_{total}$ is the initial concentration of metal ions.

## Results and discussions

Metal effect by intact cells of blue-green microalgae *Spirulina platensis* was studied as a function of metal concentration (pH 9.6). The spectrum of native *Spirulina platensis* biomass is illustrated in Fig. 1(A,B,C) (spectrum 1). Fig.1 (1) shows the absorption characteristics of control intact cells *Spirulina platensis*. The peak at 681 nm is due to the absorption of Chl a peak. At 620.7 nm is due to the absorption of phycocyanin (PC). At 500 nm is due to the absorption of carotenoides. A peak at 440 nm is due to soret band of Chl a[10]. In fig.1 (A,B,C) (2→5, 2→7) are shown also effect of Cu(II) , Cr(III) ,Cr(VI) ions on the absorption of the intact cells of cyanobacterium *Spirulina platensis*.It is seen from fig.1(A,B,C), that with increasing metal concentrations absorbtion intensity decreased for all metal ions. Data presented in fig.2 evidence the effect of initial concentrations (1mlmol) metal ions (Cu(II), Cr(III), Cr(VI) on the percent absorption intensity by *Spirulina platensis* components (carotenoides, phycocyanin, chlorophyll at 440nm,500nm, 620.7nm and 681nm wavelengths). (*Spirulina platensis* without metal ions in the picture blue color ). It is seen from fig.2, that there was a most decrease in absorption intensity from 100% (without metal ions) to 8.6 % almost 12.5-fold for Cu(II) was recorded at 500 nm, , to 13.2% at 620.7 nm , to 24.2% at 681nm and lowest decrease for Cu(II) to 34.6% at 440 nm (violet color in the picture). As for the Cr(VI), the difference is not noticeable. A most decrease in absorption intensity from 100% to 84.8% was recorded at 500nm for Cr(VI) and lowest derease to 91.2% at 440 nm (green color in the picture). Lowest effect was recorded for Cr(III): -to 96% at 500 nm and to 95% at 440 nm. ( brown color in the picture). Hence, we can assume that Cu(II) ions are most effectively associated with the active centers of components *Spirulina platensis*. As we see, although copper(II) and

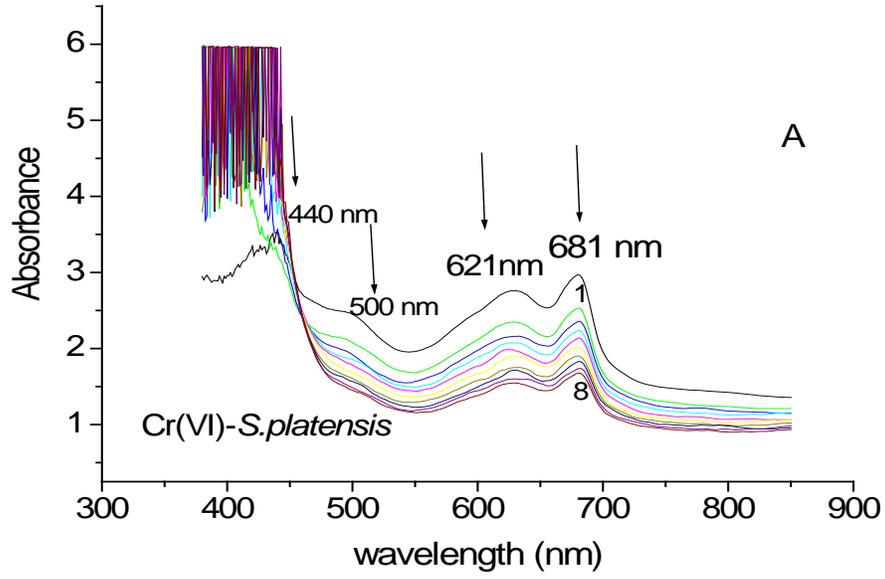

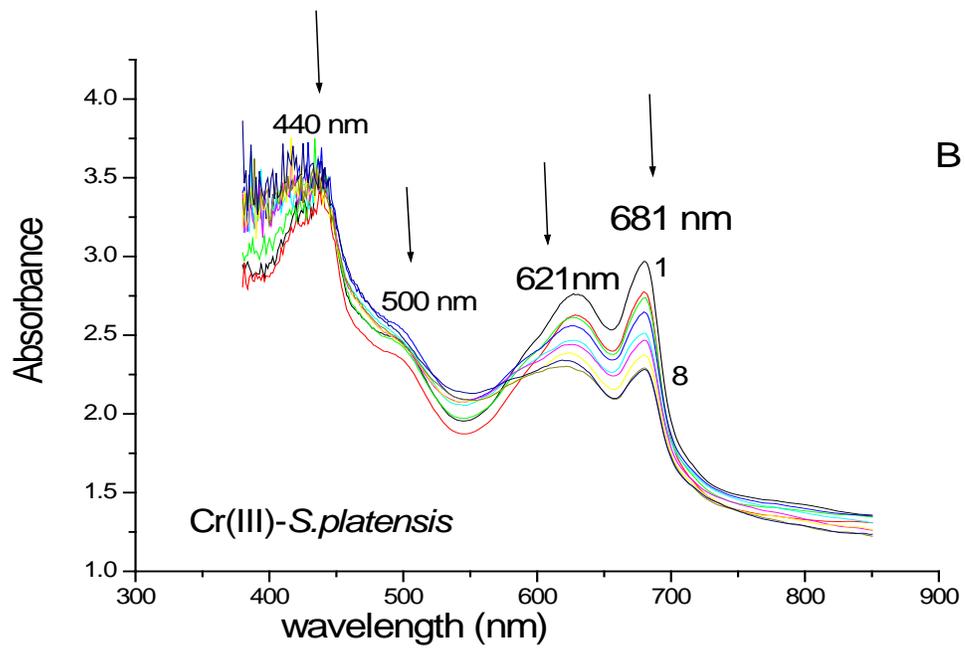

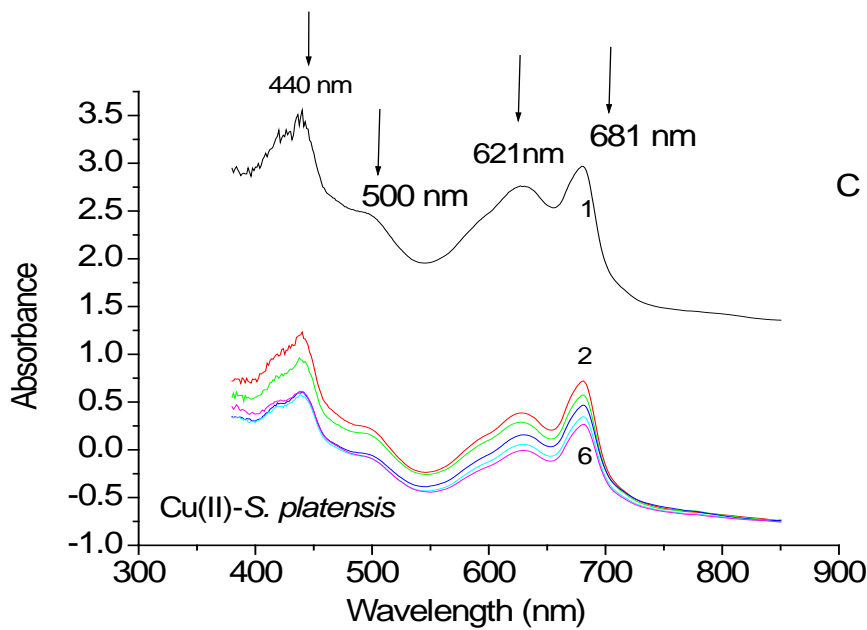

Fig.1 (A,B,C). Effect of Cr(VI), Cr(III) , Cu(II) ions on the absorption of the intact cells of the blue-green algae (cyanobacterium) *Spirulina platensis.* 1→8 [Cr(III)]=0 ÷7 mlmol; 1→8 [Cr(VI)]= 0 ÷7 mlmol; 1→6 [Cu(II)]=0 ÷5 mlmol;

chromium(VI), one cation and another anion - the trend of efficiency is the same (minimum 440 nm and maximum 500nm). *Spirulina platensis* biomass as well as that of most other marine algae[11,12] presents onto its surface net negative charge; consequently, it has low affinity for anions. Hence, Cr(VI), which is present in solution as $CrO_4^{2-}$, $HCrO_4^-$ or $Cr_2O_7^{2-}$, cannot be adsorbed onto negatively charged biomass.

In our case, the Cu(II) ions are quite effectively linked to active centers that allow us to propose, that interaction of Cu(II) ions with *Spirulina platensis* has electrostatic character, along with other specific interactions. Biosorption of ionic species on algal cell surface is fairly established process attributed to the numerous chemical groups (hydroxyl, carboxyl, carbonyl and amino groups) present on the cell surface [3]. In work [13] hypothesized that biosorption of copper by C. vulgaris takes place formation of coordination bonds between metals and amino/carbonyl groups of cell wall polysaccharides. Carboxylate groups were mainly involved in coordination and/or ionic exchange of bivalent ions, but also electron donors groups, such as amino, amide and hydroxyl groups, were likely to play a role in the biosorption process.

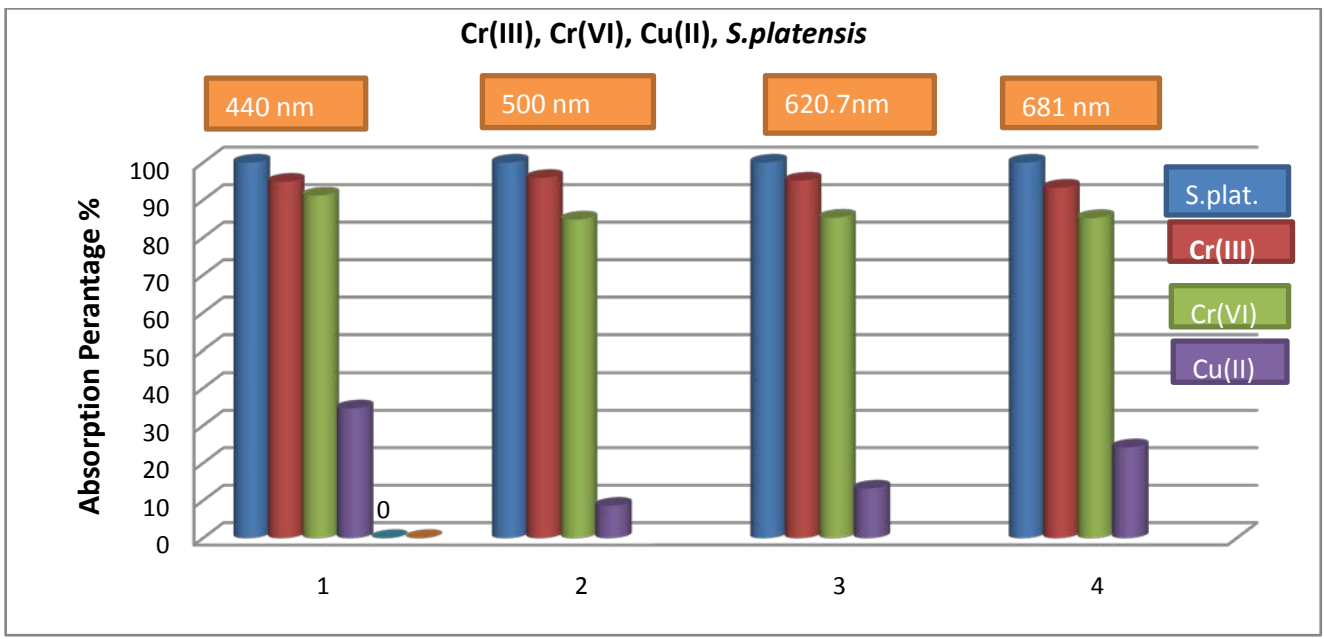

Fig.2. Effect of Cr(III), Cr(VI) and Cu(II) ions on the absorption intensity (%) at 440nm, 500 nm, 620.7 nm and 681 nm wavelengths.

As our experiments show, in contrast to copper(II), chromium (VI) is less efficient. It is known from literature [ 3], that one metal ion can occupy varying number of bonding sites and when the cell surface is surrounded by lesser number of ions, they form multiple bonds.  Which is most likely to occur in chromium (VI). As for chromium(III), this change is negligible. In water Cr(III) as aqua complex is known as octaedre complex, in which replaced  of the  water is slowly . This latter affects his biosorption.

      Biosorption of Cr(VI), Cr(III), Cu(II) ions by intact cells of  *Spirulina platensis* in medium pH 9.6, was studied as a function of metal    concentration. The linearized adsorption isotherms for these metals with *Spirulina platensis* at pH 9.6 at room temperature are shown in Fig.3 (A, B,C) by fitting experimental points.  Freundlich parameters    evaluated  from the isotherms with the correlation coefficients are given in table 1. By means of isotherms the biosorption constants ($K$) and the capacity ($n$) were determined for these complexes. In all cases, the correlation between the experimental and the theoretical data is obvious ($R$ is more than *0.9*). The results showed that the ability of the microbial cells to adsorb metal significantly varied. As seen from experimental results and from Table 1, *Spirulina platensis* showed   different binding patterns for Cu(II)_, Cr(VI)_, and Cr(III). Particularly,  significant difference between the biosorption  constants for Cu(II) –*Spirulina platensis*   and  Cr(III) _, Cr(VI)_*Spirulina platensis*  are  observed. Binding of copper(II) ions is more effective, than Cr(VI) ions and Cr(VI) more effective, than Cr(III). The adsorption capacity for Cu(II) is less (0.68), than for Cr(VI) (1.23) and Cr(III) (1.35). The biosorption constants for Cu(II)_ *Spirulina platensis*  is higher than for Cr(VI)  _  and Cr(III) _*Spirulina platensis*  2.69 and  10.15- fold  respectively. This results are in good agreement with data, which were received by us using UV-VIS spectrometry.

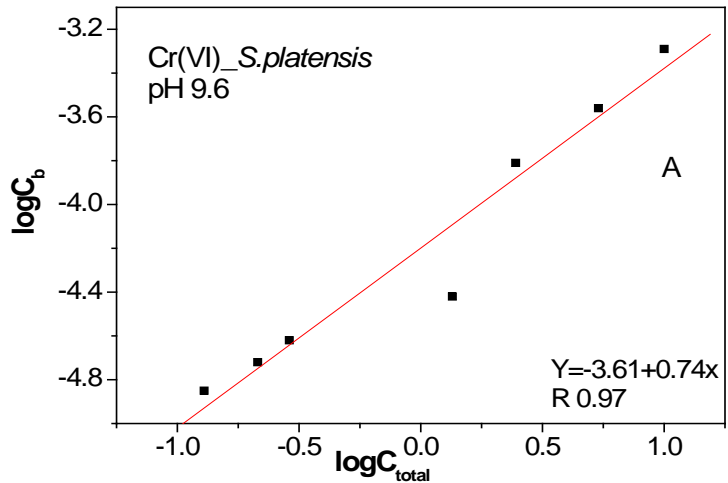

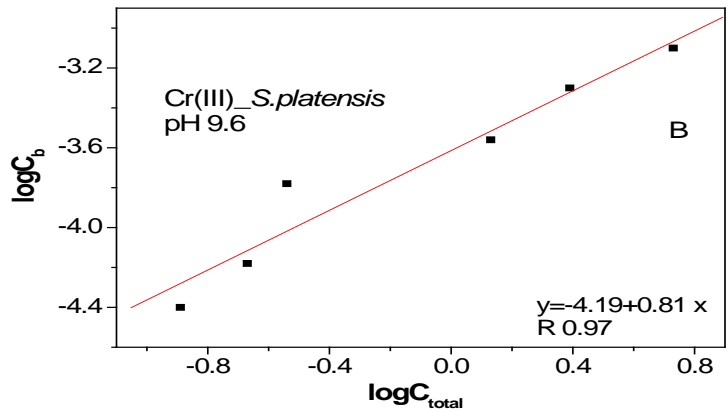

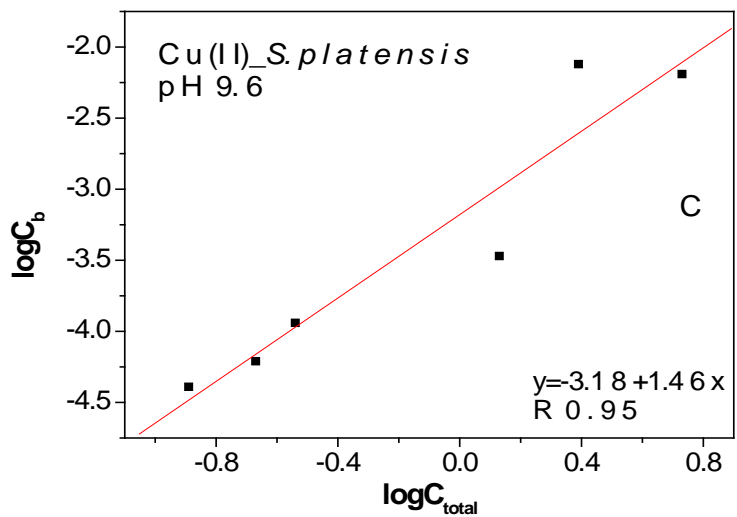

Fig. 3 (A,B,C). The linearized Freundlich adsorption isotherms for Cr(VI)_,Cr(III)_, Cu(II) - *Spirulina platensis* , where the results of biosorption experiments resulted are shown by the dots.

Table 1. Biosorption characteristics for Cr(VI)_, Cr(III)_, Cu(II) _ *Spirulina platensis* complexes in medium pH 9.6 at $23^0$ C

|  | Biosorption constant | Biosorption capacity | Correlation coefficient |
|---|---|---|---|
|  | $K \times 10^{-4}$ | n | R |
| Cu(II)-*Spirulina platensis*, pH 9.6 | 6.6 | 0.68 | 0.95 |
| Cr(III)-*Spirulina platensis* pH 9.6 | 0.65 | 1.35 | 0.97 |
| Cr(VI)-*Spirulina platensis* pH 9.6 | 2.45 | 1.23 | 0.97 |